\documentclass[conference]{IEEEtran}
\IEEEoverridecommandlockouts
\usepackage{amsmath,amsfonts,amssymb}
\usepackage{algorithm}

\usepackage{array}
\usepackage{textcomp}

\usepackage{stfloats}
\usepackage{url}
\usepackage{algpseudocode}
\usepackage{verbatim}
\usepackage{graphicx}
\usepackage{subfigure}
\usepackage{cite}
\usepackage{xspace}
\usepackage{multirow}
\usepackage[inline]{enumitem}
\usepackage[dvipsnames,svgnames]{xcolor}
\usepackage{gensymb}
\usepackage{booktabs}
\usepackage{adjustbox}
\usepackage[numbers,sort&compress,square]{natbib}
\usepackage[]{hyperref}

\usepackage[mathlines,switch]{lineno}

\def\BibTeX{{\rm B\kern-.05em{\sc i\kern-.025em b}\kern-.08em
    T\kern-.1667em\lower.7ex\hbox{E}\kern-.125emX}}

\begin{document}

\newcommand{\name}{Tempus Core\xspace}

\author{
\IEEEauthorblockN{Prabhu Vellaisamy, Harideep Nair, Thomas Kang, Yichen Ni, Haoyang Fan, Bin Qi, Jeff Chen,}
\IEEEauthorblockN{Shawn Blanton, and John Paul Shen}
\IEEEauthorblockA{\textit{ECE Department, Carnegie Mellon University} \\
\{\textit{pvellais, hpnair, thomaska, yichenn, hfan2, binq, jhchen2, rblanton, jpshen}\}\textit{@andrew.cmu.edu} \\ 
}
}


\title{\name: Area-Power Efficient Temporal-Unary Convolution Core for Low-Precision Edge DLAs}

\maketitle

\begin{abstract}

The increasing complexity of deep neural networks (DNNs) poses significant challenges for edge inference deployment due to resource and power constraints of edge devices. Recent works on unary-based matrix multiplication hardware aim to leverage data sparsity and low-precision values to enhance hardware efficiency. However, the adoption and integration of such unary hardware into commercial deep learning accelerators (DLA) remain limited due to processing element (PE) array dataflow differences. 
This work presents \textit{\name}, a convolution core with highly scalable unary-based PE array comprising of \textit{tub} (temporal-unary-binary) multipliers that seamlessly integrates with the NVDLA (NVIDIA's open-source DLA for accelerating CNNs) while maintaining dataflow compliance and boosting hardware efficiency. Analysis across various datapath granularities shows that for INT8 precision in 45nm CMOS, \name's PE cell unit (PCU) yields 59.3\% and 15.3\% reductions in area and power consumption, respectively, over NVDLA's CMAC unit. Considering a 16x16 PE array in \name, area and power improves by 75\% and 62\%, respectively, while delivering 5x and 4x iso-area throughput improvements for INT8 and INT4 precisions. Post-place and route analysis of \name's PCU shows that the 16x4 PE array for INT4 precision in 45nm CMOS requires only 0.017$\text{mm}^2$ die area and consumes only 6.2mW of total power. We demonstrate that area-power efficient unary-based hardware can be seamlessly integrated into conventional DLAs, paving the path for efficient unary hardware for edge AI inference.


\end{abstract}


\section{Introduction}

Since the introduction of AlexNet \cite{krizhevsky2012imagenet} in 2012, AI computational demands have grown rapidly \cite{wu2022sustainable}, driving the adoption of hardware accelerators like GPUs and TPUs for deep learning (DL) training and inference. However, DL's computational needs have quickly outpaced the growth of available compute power, leading to what is now termed the ``AI compute gap". This gap has renewed interest in innovative computational techniques and architectures aimed at closing this compute performance and energy efficiency gap.

\begin{figure}[t!]
\centering
\includegraphics[width=0.475\textwidth, height = 5cm]{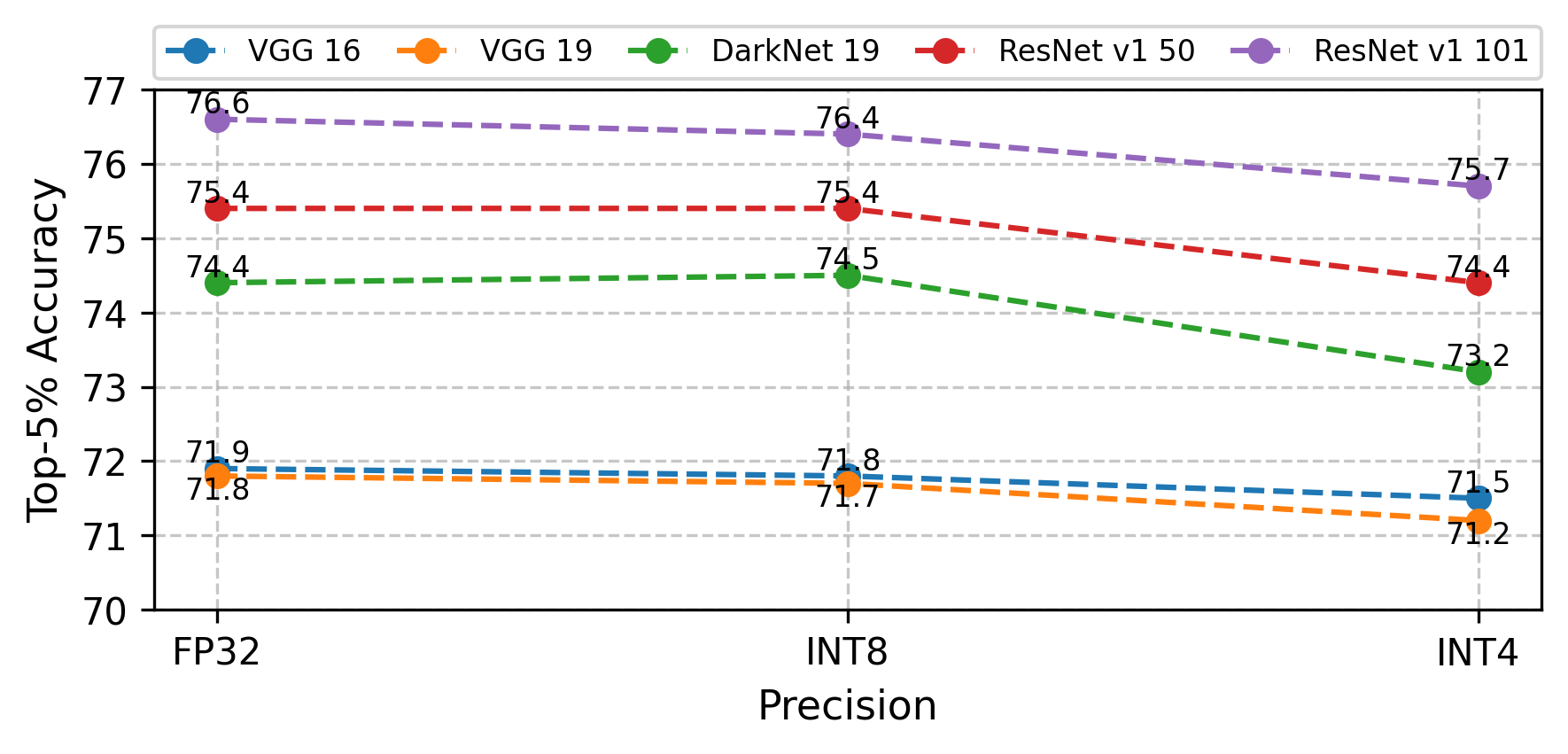}
\caption{Quantization training accuracies achieved on different ImageNet CNNs for different integer-based precisions when compared to baseline FP32 precision \cite{jain2020trained}. Results show minimal accuracy decrease with lower precisions.}

\label{fig:quant}
\end{figure}

In recent years, quantization techniques have become an effective strategy for deploying AI models with reduced model complexity. While 32-bit floating point (FP32) was the standard for DL training, FP16 is now widely adopted, offering 4x-8x performance improvements \cite{sun2020ultra}. Building on this trend, FP8 precision for both training and inference has been proposed \cite{micikevicius2022fp8} and implemented by Nvidia in their H100 GPUs, which support FP8 models through the Transformer Engine library \cite{choquette2022nvidia}. In particular, FP8 training on various convolutional neural network (CNN) models has shown minimal accuracy degradation while increasing throughput by 2x-4x \cite{sun2020ultra}. In addition to floating point formats, 8-bit integer (INT8) training has been demonstrated to reduce the training time of CNNs on Pascal GPUs by 22\% \cite{zhu2020towards}. Although INT8 has become the standard for DL inference, a shift toward INT4 precision has led to a 77\% performance improvement over INT8 \cite{han2020convolutional}. The results of ImageNet trainings for quantized CNN models and their corresponding Top-5\% accuracy are presented in Fig. \ref{fig:quant}, showing minimal accuracy degradation for INT4 quantized models compared to their baseline FP32 implementations. These quantization results offer tremendous promise particularly for edge inference deployment, where computational resources are limited.

Unary computing has recently emerged as a promising paradigm for low-precision AI, enabling highly area- and power-efficient hardware by trading off latency for efficiency. In particular, temporal encoding techniques have been shown to significantly reduce computational overhead in matrix multiplication units, as demonstrated by recent works \cite{nair2023tugemm} \cite{vellaisamy2023tubgemm} \cite{vellaisamy_isvlsi24}, which exploit the inherent sparsity of DL models. However, these existing implementations focus primarily on GEMM-level designs and are not scaled to inherently support convolution dataflows. This work addresses that gap by extending unary computing to convolution dataflows, while ensuring compatibility with widely-used deep learning accelerators (DLAs) like NVDLA \cite{nvdla}. Table \ref{tab:sparsity} illustrates the inherent weight sparsity for INT8 quantized CNNs that can be exploited by unary computing.

\begin{table}[t!]
    \caption{Word sparsity indicating the percentage of zero weights in eight quantized INT8 CNNs \cite{vellaisamyexploration}}
    \label{tab:sparsity}
    \centering
    \begin{adjustbox}{max width=\linewidth}
    \begin{tabular}{c|c}
        \toprule
        \textbf{CNN} & \textbf{Word (\%) 8 bits}   \\
        \midrule
        \midrule
        MobileNetV2 & 2.25 \\
        \midrule
        MobileNetV3 & 9.52 \\
        \midrule
        GoogleNet & 1.91 \\
        \midrule
        InceptionV3 & 1.99 \\
        \midrule
        ShuffleNetV3 & 1.43 \\
        \midrule
        ResNet18 & 2.043 \\
        \midrule
        ResNet50 & 2.45 \\
        \midrule
        ResNeXt101 & 2.64 \\
        \bottomrule
    \end{tabular}
    \end{adjustbox}
\end{table}

Despite recent progress, there remains a substantial gap in research. Only a few custom unary-based accelerators have been proposed, such as in \cite{li2018scope} and \cite{thakkar2023low}. These accelerators rely on stochastic computation, which trades off computational accuracy for hardware efficiency. However, as lower precisions become the standard, deterministic computation becomes essential to minimize accuracy degradation. Furthermore, these stochastic designs are not easily compatible with existing off-the-shelf DLAs, which are widely used in today's installed hardware base.

This work presents \textit{\name}, a highly hardware-efficient temporal-unary-binary convolution engine designed to integrate seamlessly into existing DLAs without sacrificing compute accuracy. \textit{\name} employs \textit{tub} multipliers \cite{vellaisamy_isvlsi24} within its processing element (PE) array to execute convolution operations. Seamless integration into current DLAs is demonstrated through its incorporation into the NVDLA architecture \cite{nvdla}. While maintaining the computational accuracy of binary-based arithmetic designs, \textit{\name} delivers improved hardware efficiency compared to the NVDLA convolution core and remains fully compatible with NVDLA’s dataflow.

The main contributions of this work are as follows.

\begin{enumerate}
    \item \textit{Novel High Iso-Area Throughput Convolution Engine}: \name employs \textit{tub} multipliers for unary-based convolution, designed as a drop-in replacement for modern DLA convolution datapaths. Unlike previous temporal GEMM designs \cite{nair2023tugemm}\cite{vellaisamy2023tubgemm} that follow an outer-product GEMM dataflow, \name serves as a convolution engine supporting inner-product convolution dataflow, improving efficiency without sacrificing compatibility. For \name comprising of $16 \times 16$ PE array, 5x and 4x iso-area throughput is realized for INT8 and INT4 precisions, compared to standard binary-based NVDLA.
    

    \item \textit{Integration with Industry-Grade DLAs}: Unlike previous unary DLA datapaths, \name is designed for seamless integration into existing binary-based DLAs, enhancing key hardware performance and efficiency metrics while maintaining functional integrity. This integration enables programmers to retain the full functionality of the DLA’s existing software stack. 
    
    \item\textit{Comprehensive Comparative Evaluation}: We perform a detailed comparison between the convolution datapath of NVDLA’s convolution core (CC) and \name across multiple levels of hierarchy, ranging from a single ``PE cell" to PE array (multiple cells), and up to entire PCU (PE Cell Unit). Rigorous evaluation is performed across low integer precisions and array sizes using 45nm CMOS technology to report post-synthesis metrics.

    \item \textit{Place-and-Route Analysis}: Unlike previous studies, we conduct a post place-and-route analysis to obtain more precise measurements of area and power between a \name's PE cell unit (PCU) and NVDLA's CMAC unit for $16 \times 4$ array. Results show improvements of 53\% in area efficiency and 44\% in power efficiency, respectively, in the case of \name's PCU.

    \item \textit{Fine-Grained CNN Profiling}: Additionally, weight distribution analysis for PE array tile size of 16x16 is conducted across convolution layers for INT8-quantized MobileNetV2 and ResNeXt101 models. This analysis provides insights into application-specific, sparsity-driven latency and thereby energy for \name.
\end{enumerate}

This paper is structured as follows. Section \ref{sec:background} provides the background and overview of relevant topics. Section \ref{sec:micro} describes the \name design and its integration into the NVDLA design. Section \ref{sec:evaluation} discusses the evaluation methodology used in this study, with Section \ref{sec:results} reporting the results of the experiments along with their analysis. Finally, Section \ref{sec:conc} summarizes conclusions and future work. 

\section{Background and Related Works}
\label{sec:background}

\subsection{General Matrix Multiplication}

General matrix multiplication (GEMM) is a key DL operation involving activation and weight matrix multiplication, defined in the BLAS library \cite{blackford2002updated}, and can be expressed as:

\begin{equation}
    O:= {\alpha}A{\times}B+{\beta}C
\end{equation}

where $O$ is the output matrix of size $M{\times}P$, and $A$, $B$, and $C$, are input matrices of sizes $M{\times}N$, $N{\times}P$, $M{\times}P$, respectively. $\alpha$ and $\beta$ are scaling factors. 

The compute-intensive nature of fully connected and convolutional layers allows them to be efficiently mapped to matrix multiply units. A study indicates that up to 89\% of CPU runtime and 95\% of GPU runtime are spent processing fully connected and convolutional layers in AlexNet \cite{jia2014learning}, underscoring the critical role of GEMM in DL computations.

\subsection{\textit{tub}GEMM}

\begin{figure}[h]
\centering
\includegraphics[width=0.27\textwidth, height=4.2cm]{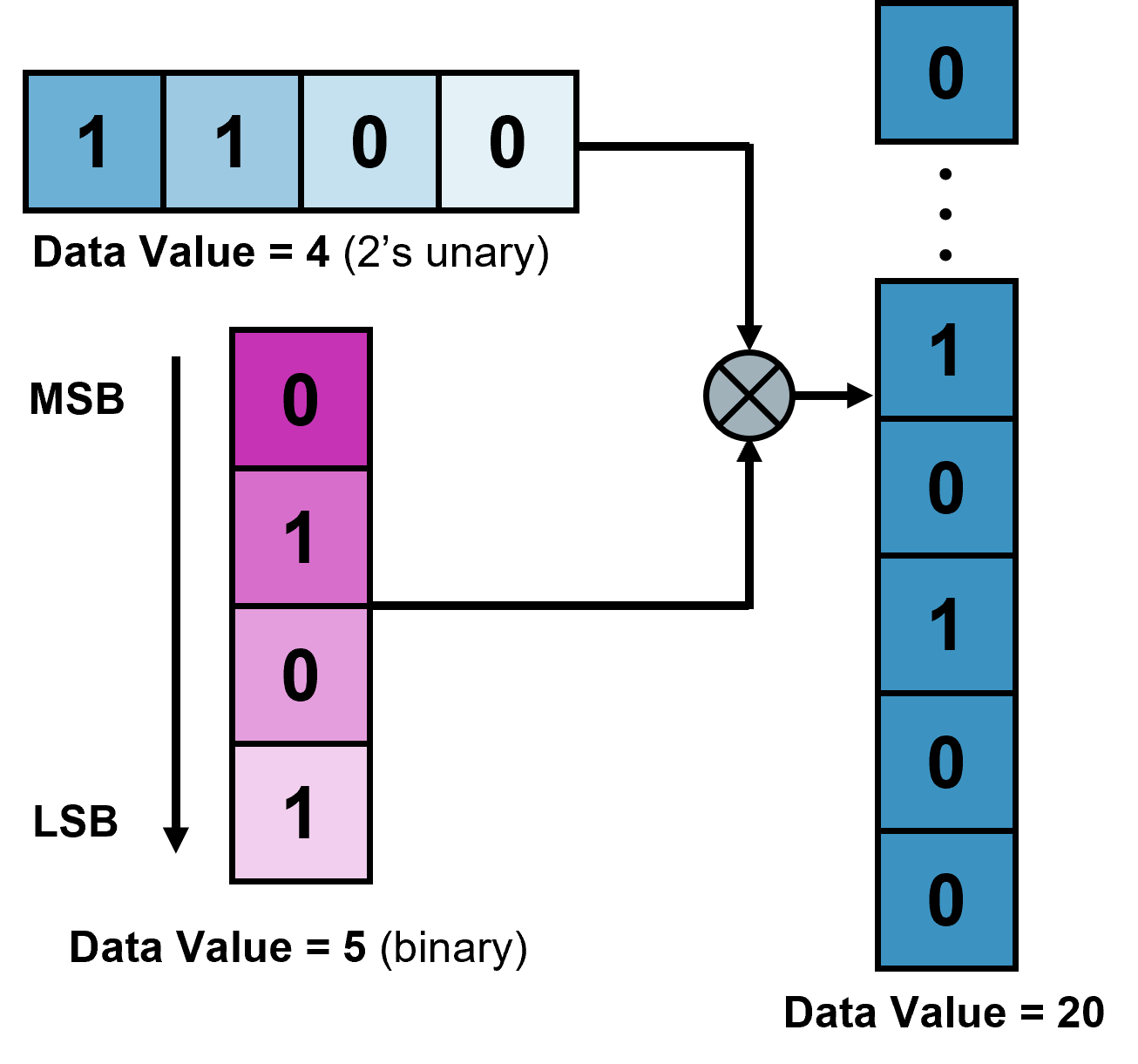}
\caption{An example dataflow of an INT4 \textit{tub} multiplier. The INT4 \textit{tub} multiplier take a 4-bit binary-encoded value and a single temporal-coded bitstream as inputs. For each ``1'' bit in the bit-serial temporal-coded input, the binary value is accumulated, producing the desired output result.}
\label{fig:tub}
\end{figure}

\textit{tub}GEMM \cite{vellaisamy2023tubgemm}\cite{vellaisamy_isvlsi24} is a novel temporal-binary hybrid INT-based GEMM architecture that inherently leverages the sparsity found in DL workloads, resulting in significant improvements in latency and energy efficiency. This architecture processes activation data as binary inputs while encoding weight bits into a single temporal bitstream. The processing element (PE) of \textit{tub}GEMM consists of multiplexers, shifters, and registers, contributing to a streamlined microarchitectural design. It employs a unique \textit{2s}-unary encoding scheme, where each unary bit or cycle is interpreted as a data value of 2, effectively halving the latency.  This advancement significantly improves upon the performance of \textit{tu}GEMM \cite{nair2023tugemm}, which has a worst-case latency of $N*{(2^{w-1})}^2$ cycles. In contrast, \textit{tub}GEMM reduces the worst-case latency to $N*{(2^{w-2})}$ cycles, where $N$ represents the common dimension of input matrices, and $w$ denotes the bitwidth. Fig. \ref{fig:tub} illustrates the dataflow of an INT4 \textit{tub} multiplier. A comparative design analysis of existing unary-based GEMM architectures \cite{vellaisamy_isvlsi24} concludes \textit{tub}GEMM to be the most optimal unary design. 


\subsection{NVDLA Accelerator}

 NVDLA \cite{nvdla} is an open-source, industry-grade inference engine developed by NVIDIA, integrated into Xavier systems on chip (SoCs) for self-driving applications. It offers an end-to-end software-hardware stack that provides a DL framework along with a runtime environment and spans all the way to RTL implementation. This study employs the \textit{nv\_small} configuration, which is tailored for embedded and cost-sensitive applications \cite{nvdla}. Unlike the \textit{nv\_large} variant, the \textit{nv\_small} design does not include a dedicated control processor; instead, it relies on the host processor for task scheduling, memory management, and NVDLA control. Although other open-source DLAs, such as Gemmini \cite{genc2021gemmini}, exist, they are primarily academic designs focused on GEMM acceleration. Moreover, NVDLA was reported to be 3.77x faster than Gemmini when comparing equivalently sized configurations running ResNet-50 \cite{gonzalez2020chipyard}. Given its performance and proven industry reliability, we select NVDLA as our experimental platform.

 \begin{figure}[t!]
\centering
\includegraphics[width=0.48\textwidth, height = 14 cm ]{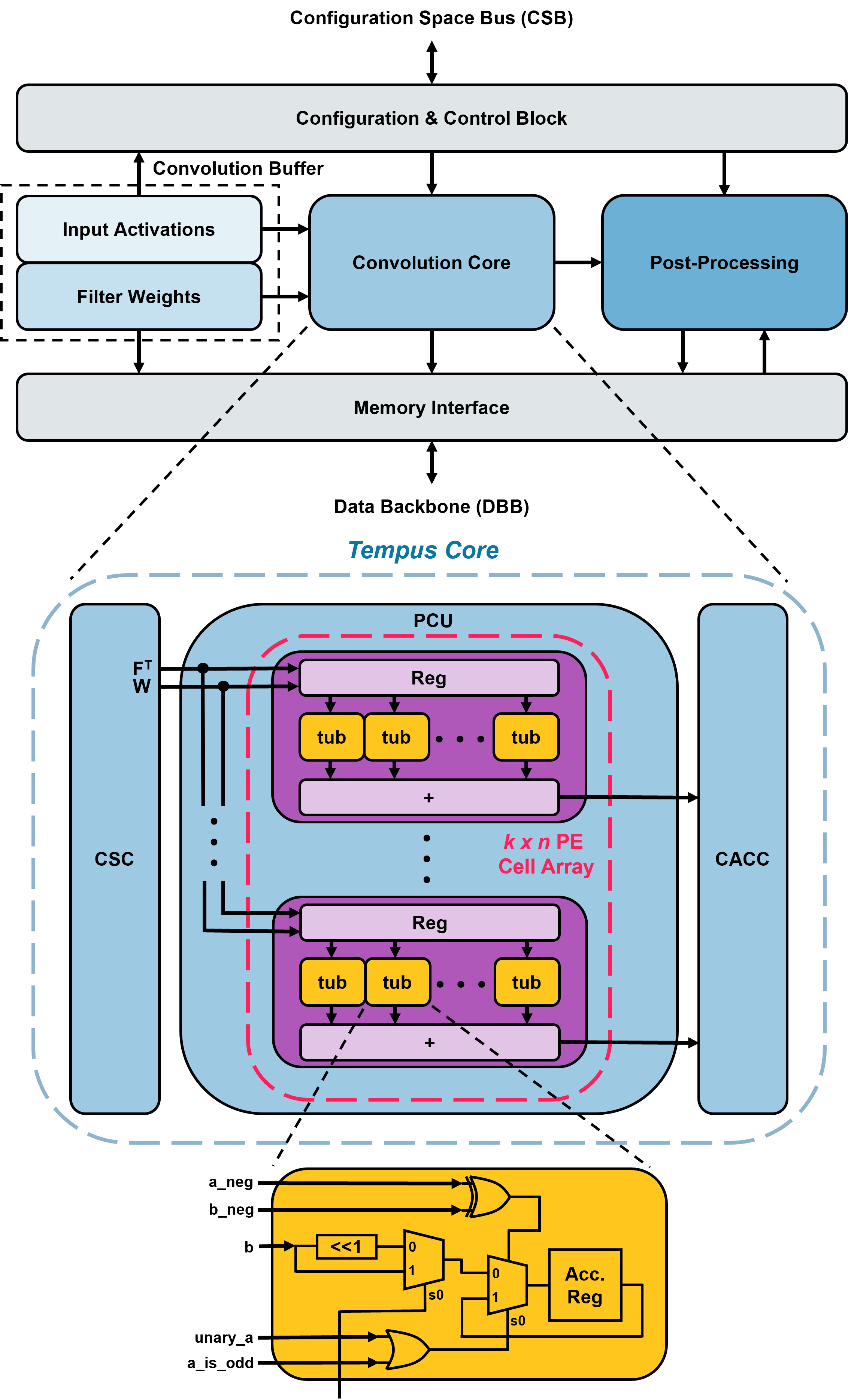}
\caption{Overview of \name integration into NVDLA. In NVDLA, the convolution buffer (CB) stores both activation and weight values, which are fed into the Convolution Core (CC) consisting of the convolution sequence controller (CSC), the Convolution MAC (CMAC) unit (containing the $k \times n$ MAC array, output registers, and handshaking logic) and the convolution accumulator (CACC). In this work, CC is replaced by \name, which contains modified CSC and a PE cell unit (PCU) containing $k$ \textit{tub}-based PE cells replacing CMAC. Each PE cell consists of $n$ \textit{tub}-based multipliers. Additional handshaking logic to facilitate multi-cycle convolution operation is also present, as well as output registers to maintain functionality.}

\label{fig:core}
\end{figure}
 
 Fig. \ref{fig:core} provides a high-level overview of NVDLA's convolution pipeline. It comprises (i) the convolution buffer (CB), which stores input activations and filter weights, (ii) the convolution core (CC), which serves as the convolution datapath, and (iii) the post-processing unit, which includes the activation engine, the pooling engine, and other components essential for accelerating CNNs. The primary datapath performing MAC operations to calculate partial sums in CC is the convolution MAC (CMAC) unit. The CMAC unit consists of $k$ processing element (PE) cells (termed MAC Cells in \cite{nvdla}), each with $n$ multipliers, producing a partial result for each kernel. Data is broadcast from the CB via a convolution scheduler (CSC) to the \textit{k} PE cells, with partial results accumulated in the adder trees of the convolution accumulation (CACC) unit. Each cell supports clock gating to reduce dynamic power consumption during idle or underutilized conditions when the kernel count is insufficient for full utilization. In addition, CMAC features intermediate registers that facilitate retiming and pipelining.

 To summarize the NVDLA microarchitecture hierarchy:
 \begin{enumerate}
     \item The convolution engine in NVDLA is called convolution core (CC), the focus of this work.
     \item The CC comprises CSC, CMAC, and CACC. The CMAC unit contains $k \times n$ PE array, denoting $k$ number of PE (MAC) cells, each consisting $n$ multipliers and registers for caching partial sums.
     \item In addition to the $n$ multipliers, each PE cell contains local registers and an adder tree to accumulate the intermediate results, producing one partial sum corresponding to the PE cell. 
 \end{enumerate}

\section{\name Microarchitecture}
\label{sec:micro}
The \textit{\name} microarchitecture (Fig. \ref{fig:core}) is implemented as a temporal-unary-binary (\textit{tub}) hybrid convolution engine, designed as a drop-in replacement for the convolution core (CC) in NVDLA. \name adheres to the original dataflow in NVDLA and can directly replace its convolution core. \name consists of a modified CSC, PE cell unit (PCU), and the CACC. The PCU is analogous to the CMAC unit in NVDLA, containing $k \times n$ \textit{tub}-based PE array for efficient MAC operation. In this setup, direct convolution involves sharing the feature data array across $k$ PE cells, with each cell caching a different weight data array. Each PE cell contributes a partial sum, producing $k$ partial sums per feature and weight data array. These partial sums are subsequently accumulated in the CACC unit. While NVDLA's CMAC unit generates $k$ partial sums per cycle, the \name'sPCU generates $k$ partial sums over multiple cycles, with the cycle count equal to the largest weight magnitude in the $k\times n$ array. The modified CSC in \name feeds transposed feature data since $W \times F^T = \text{accum}(W \odot F)$. Note each PE cell also incorporates $n$ \textit{2s}-unary blocks into the temporal encoder, allowing optimal temporal-to-binary data conversion. The registers corresponding to the PE cells cache the partial sums, which are only forwarded to the CACC once all partial sums have been generated across the cells.

Having detailed the key components of \name, we now outline its dataflow structure and its seamless integration into existing DLAs.
\begin{itemize}
    \item The PCU comprises a $k \times n$ PE array, with dedicated registers for caching outputs and handshaking logic to coordinate multi-cycle operations.
    \item The input data and initial weight data cubes are divided into $1{\times}1{\times}n$ element cubes. Each PE cell caches one weight cube from the weight kernel, and the single input data cube is shared between the $k$ PE cells.
    \item Each \textit{tub} multiplier performs a multiplication between a single weight and input data element over multiple cycles, and the adder tree in the cell accumulates the results into a partial sum to be fed next to the CACC unit.
\end{itemize}

Further microarchitectural enhancements in \textit{\name} include the integration of additional handshaking protocols with buffer blocks to accommodate multiple \textit{tub} cycles per partial sum computation within each PE cell. This modification ensures efficient synchronization and data integrity. The number of compute cycles for $k \times n$ array in \textit{\name} is determined by the largest weight magnitude present in the array. We profile quantized CNNs to report application-specific latency per array in Section \ref{sec:results}.

\section{Evaluation Methodology}
\label{sec:evaluation}

This section details the evaluation setup, profiling methodology, and key EDA tools used to measure latency and energy values across designs. Comparative analysis between NVDLA's CC and \name is performed at three levels of granularity: (i) single PE cell, (ii) $k \times n$ PE array, and (iii) entire CMAC unit versus PCU in \name. The analysis spans both INT8 and INT4 precisions, with varying $k \times n$ configurations. The designs were synthesized using NanGate45 open-source cell library in Synopsys Design Compiler, operating at a fixed 250 MHz clock frequency to maintain consistent timing across evaluations. The binary PE cells are elaborated with DesignWare-optimized multipliers and Wallace adder tree during synthesis. Furthermore, place-and-route results are obtained using Cadence Innovus with the NanGate45 library to further extend the evaluation.
 
 Weight-value profiling is performed on two INT8-quantized models, MobileNetV2 and ResNeXt101, to determine application-specific latency and energy consumption. Using a $16 \times 16$ ($k$=16, $n$=16) max pool across weights present in the model's convolution layers, the largest weight value within each 16x16 tile is determined and its frequency of occurrence as the largest value across is derived. This directly correlates to the compute cycles since the largest value across an array of 16 PE cells with 16 multipliers bottlenecks the \textit{tub} model compute. Note the PCU takes a few extra cycles for caching in and out the values, hence analysis pertains to the array. Additionally, sparsity is analyzed in a similar fashion to estimate the average number of ``silent" PEs per array, where \textit{tub} multipliers remain inactive for zero-valued weights.

\section{Results}

\label{sec:results}

\begin{table}
\caption{Post-synthesis cell area (Top) and total power (bottom) results in 45nm CMOS: Single PE Cell ($k=1)$}
\label{tab:pe_area}
\centering
\begin{adjustbox}{max width=\linewidth}
\begin{tabular}{|cc|c|c|l|}
\hline
\multicolumn{2}{|c|}{\textbf{Configuration}}                       & \multirow{2}{*}{\textbf{\begin{tabular}[c]{@{}c@{}}Binary PE Cell \\ Area ($\mu$m$^2$)\end{tabular}}} & \multirow{2}{*}{\textbf{\begin{tabular}[c]{@{}c@{}}\textit{tub} PE Cell\\ Area ($\mu$m$^2$)\end{tabular}}} & \multicolumn{1}{c|}{\multirow{2}{*}{\textbf{\begin{tabular}[c]{@{}c@{}}Improvement \\ (\%)\end{tabular}}}} \\ \cline{1-2}
\multicolumn{1}{|c|}{\textbf{Precision}}    & \textbf{Number of PEs ($n$)} &                                                                                               &                                                                                          & \multicolumn{1}{c|}{}                                                                                      \\ \hline
\multicolumn{1}{|c|}{\multirow{3}{*}{INT4}} & 16                   &  0.0022                                                                                            &   0.0006                                                                                     &   \multicolumn{1}{|c|}{71.89}                                                                                                         \\ \cline{2-5} 
\multicolumn{1}{|c|}{}                      & 256                  &    0.0371                                                                                         & 0.0046                                                                                      &  \multicolumn{1}{|c|}{87.53}                                                                                                          \\ \cline{2-5} 
\multicolumn{1}{|c|}{}                      & 1024                 &  0.1462                                                                                            & 0.0171                                                                                       &   \multicolumn{1}{|c|}{88.30}                                                                                                         \\ \hline \hline
\multicolumn{1}{|c|}{\multirow{3}{*}{INT8}} & 16                 &  {0.0056}                                                                                          &    {0.0011 }                                                                                   &  \multicolumn{1}{|c|}{{80.15}}                                                                                                          \\ \cline{2-5} 
\multicolumn{1}{|c|}{}                      & 256                  & {0.1063}                                                                                             &   {0.0093}                                                                                      &  \multicolumn{1}{|c|}{ {91.24}}                                                                                                          \\ \cline{2-5} 
\multicolumn{1}{|c|}{}                      &  1024               &  {0.4334}                                                                                         &  {0.0355}                                                                                    &     \multicolumn{1}{|c|}{ {91.81}}                                                                                                       \\ \hline
\end{tabular}
\end{adjustbox}
\end{table}


\begin{table}

\begin{adjustbox}{max width=\linewidth}
\begin{tabular}{|cc|c|c|l|}
\hline
\multicolumn{2}{|c|}{\textbf{Configuration}}                       & \multirow{2}{*}{\textbf{\begin{tabular}[c]{@{}c@{}}Binary PE Cell \\ Power (mW)\end{tabular}}} & \multirow{2}{*}{\textbf{\begin{tabular}[c]{@{}c@{}}\textit{tub} PE Cell\\ Power (mW)\end{tabular}}} & \multicolumn{1}{c|}{\multirow{2}{*}{\textbf{\begin{tabular}[c]{@{}c@{}}Improvement \\ (\%)\end{tabular}}}} \\ \cline{1-2}
\multicolumn{1}{|c|}{\textbf{Precision}}    & \textbf{Number of PEs ($n$)} &                                                                                               &                                                                                          & \multicolumn{1}{c|}{}                                                                                      \\ \hline
\multicolumn{1}{|c|}{\multirow{3}{*}{INT4}} & 16                  &   0.09                                                                                             &   0.06                                                                                      & \multicolumn{1}{|c|}{25.86}                                                                                                           \\ \cline{2-5} 
\multicolumn{1}{|c|}{}                      & 256                  &    1.03                                                                                           &  0.19                                                                                       & \multicolumn{1}{|c|}{81.74}                                                                                                           \\ \cline{2-5} 
\multicolumn{1}{|c|}{}                      & 1024                 &  3.98                                                                                             & 0.51                                                                                         &   \multicolumn{1}{|c|}{87.25}                                                                                                         \\ \hline \hline
\multicolumn{1}{|c|}{\multirow{3}{*}{INT8}} & 16                   &   {0.20}                                                                                            &     {0.088}                                                                                      &   \multicolumn{1}{|c|}{ {54.72}}                                                                                                         \\ \cline{2-5} 
\multicolumn{1}{|c|}{}                      &  256                 & {3.00}                                                                                              &  {0.32}                                                                                       &    \multicolumn{1}{|c|}{ {89.35}}                                                                                                        \\ \cline{2-5} 
\multicolumn{1}{|c|}{}                      &  1024                &  {12.20}                                                                                          &  {1.06}                                                                                      &  \multicolumn{1}{|c|}{91.28}                                                                                                          \\ \hline
\end{tabular}
\end{adjustbox}

\end{table}

\begin{figure}
\centering
\includegraphics[width=0.5\textwidth, height= 4.4cm]{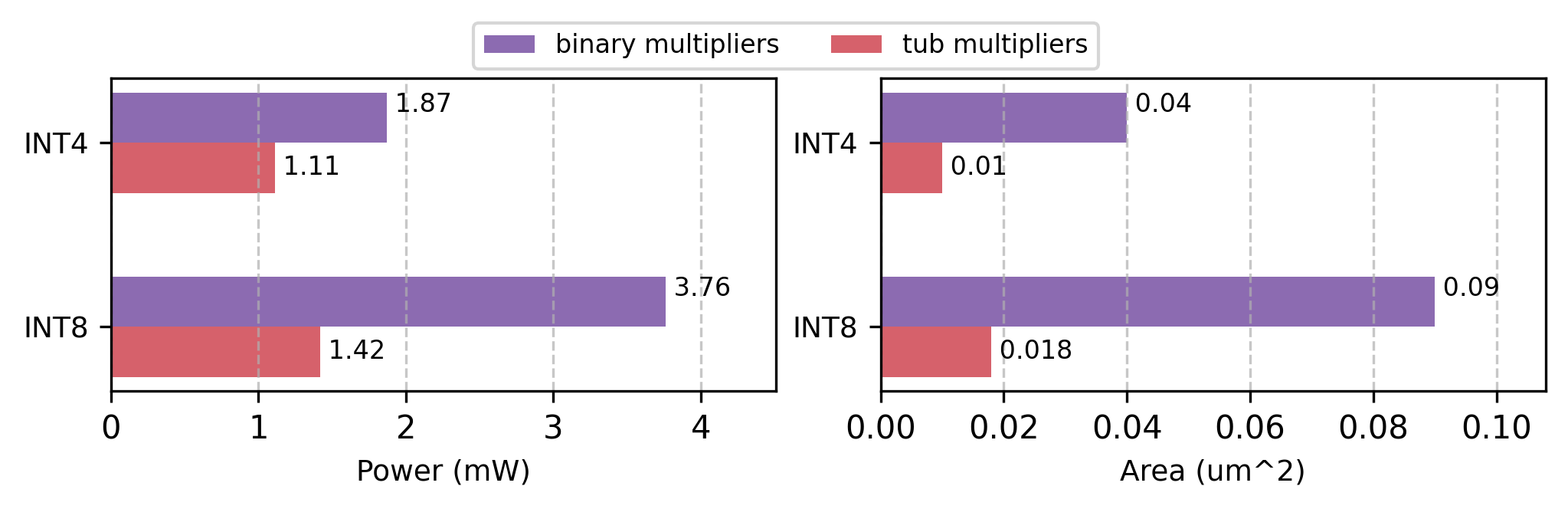}
\caption{Post-synthesis total power consumption (left) and cell area utilization (right) in 45nm CMOS for the two different $16 \times 16$ designs, both for INT4 and INT8 precisions.}
\label{fig:pe_16_area_power}
\end{figure}

\subsection{Area-Power Efficiency}

Post-synthesis results for a single PE cell, containing registers, multipliers, and an adder tree, are summarized in Table \ref{tab:pe_area}. Single \textit{tub}-based PE cells significantly outperform their binary counterparts in both area and power efficiency. For INT8, tub multipliers reduce area by 91.8\% and power by 91.3\% for n=1024 PEs. INT4 implementations achieve 88.3\% area and 87.2\% power reductions. PE cell with \textit{tub} multipliers demonstrate superior scalability across $n$: area scales by 7.7x for INT4 and 8.5x for INT8, compared to 16.9x and 19x for binary PE cell. Power consumption scales by 2.9x for INT4 and 3.5x for INT8, versus 11.4x and 15x for binary PE cell.

\name’s scalability is maintained as we scale from a single PE cell to a $16 \times 16$ array. At INT8 precision, the binary-based implementation requires 0.09 ${\mu}\text{m}^2$ of area and 3.8 mW of power, respectively. In comparison, \name's \textit{tub}-based PE cells consume only 0.018 ${\mu}\text{m}^2$ and 1.42 mW, achieving a 75\% area reduction and 62\% power savings. Similarly, for INT4, the reductions are 80\% in area and 41\% in power. These results are illustrated in Fig. \ref{fig:pe_16_area_power}.

\begin{figure}
\centering
\includegraphics[width=0.48\textwidth, height= 4.1cm]{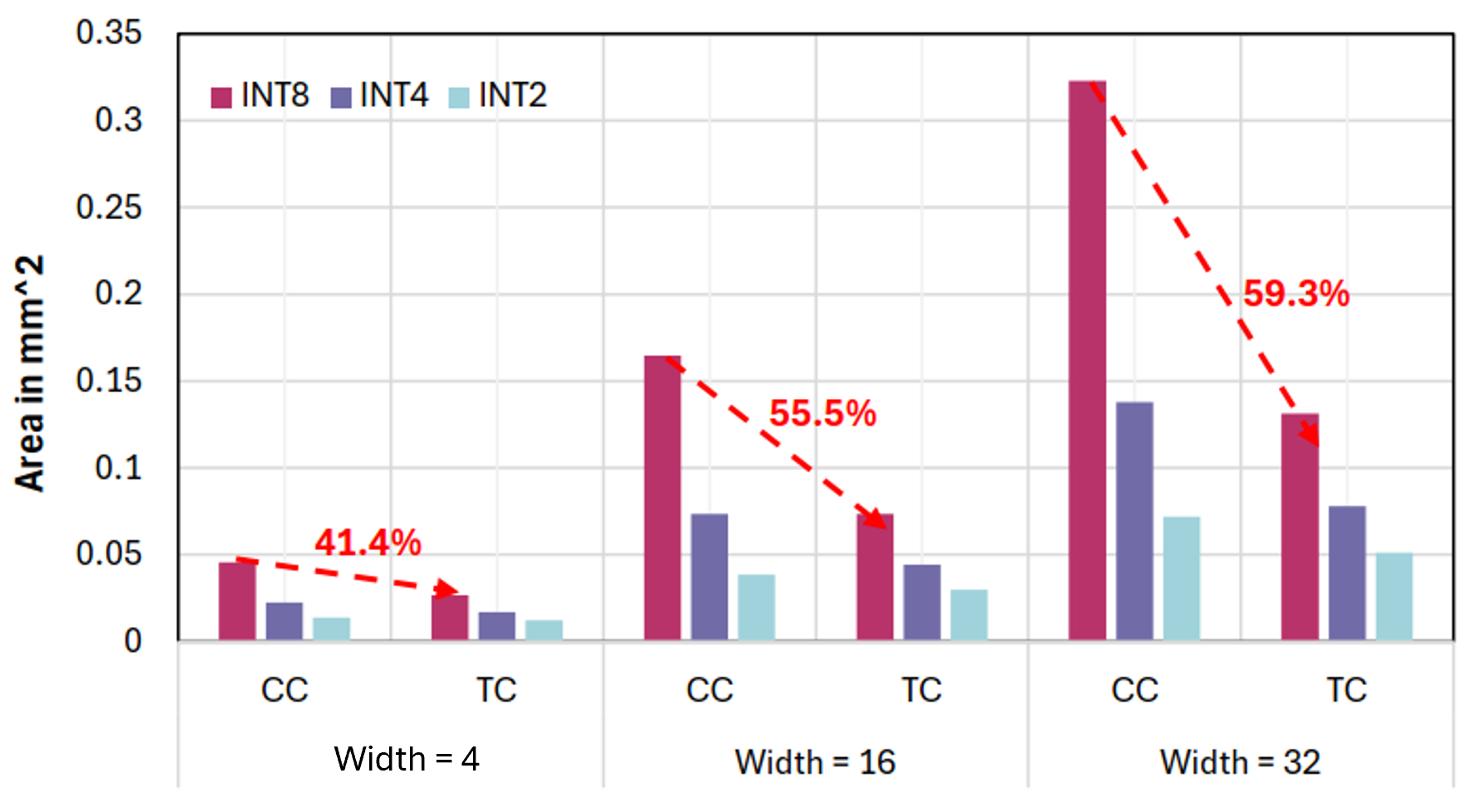}
\includegraphics[width=0.48\textwidth, height= 4.1cm]{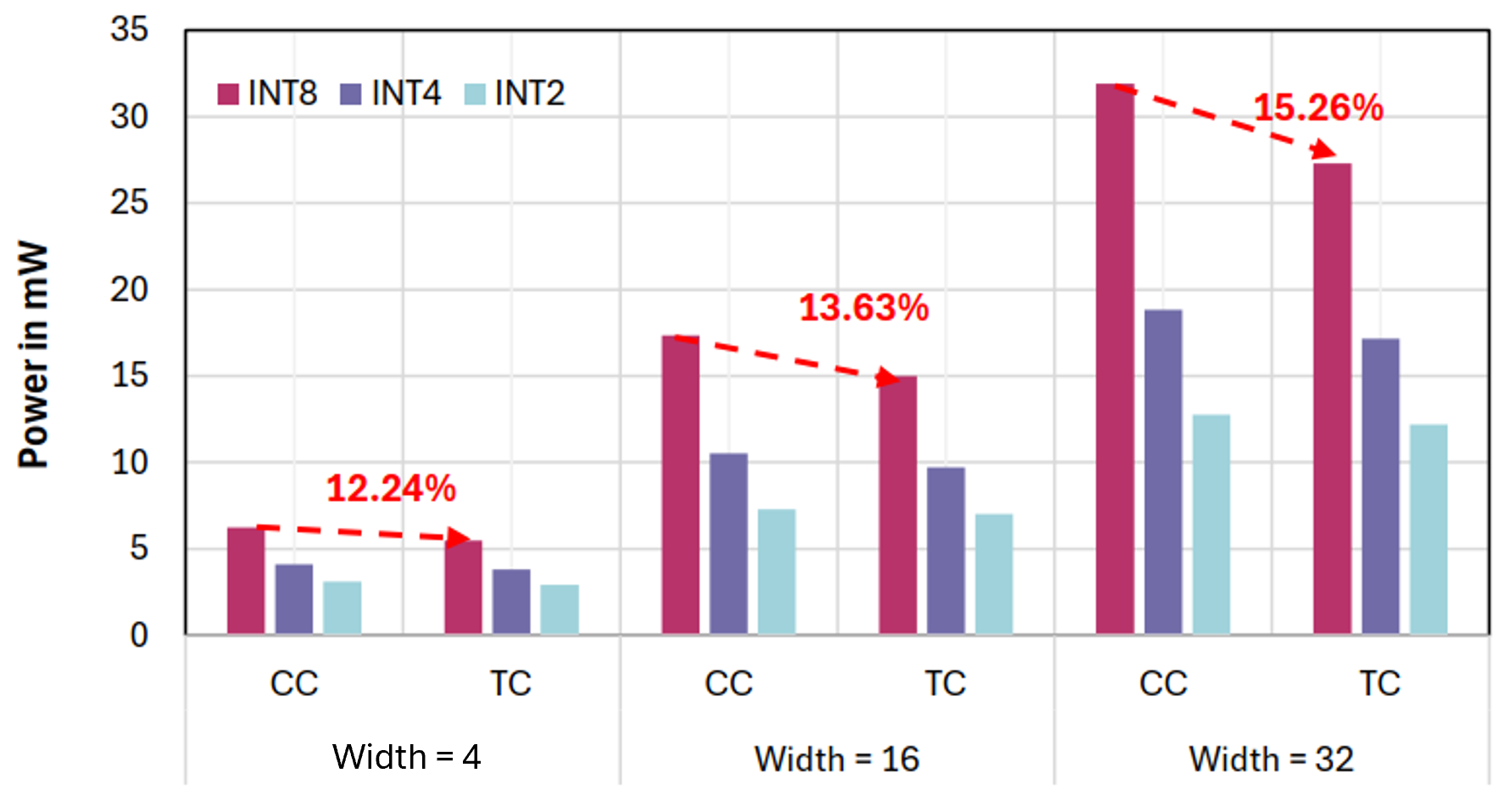}
\caption{Post-synthesis area utilization and total power consumption in 45nm CMOS across entire CMAC and PCU units for different array widths ($16 \times n$) with n = 4, 16, and 32. CC refers to the CMAC unit in CC, and TC denotes the PCU inside \name. For a constant core configuration of 16 PE Cells (array height), number of multipliers are varied across INT8, INT4, and INT2 precisions. Percentage decrease in area and power consumption for INT8 are denoted by the red dotted arrows.}
\label{fig:core_eval}
\end{figure}

Finally, moving further up in the hierarchy, post-synthesis results of the entire CMAC unit and the PCU of \name are compared and illustrated in Fig. \ref{fig:core_eval}, for various widths ($n$) in INT8, INT4, and INT2 precisions. The PCU improves area and power consumption by 59.3\% and 15.3\%, respectively.


\textbf{Key Takeaway:} PCU (and in turn, \name) demonstrates significantly superior area-power efficiency compared to the baseline binary-based CMAC unit across diverse integer-based precisions and granularities of compute datapath, paving the path for more silicon-optimized and scalable DLAs.

\subsection{Place-and-Route Results}

\begin{figure}[t!]
    \begin{subfigure}
        \centering
        \includegraphics[height=1.4in]{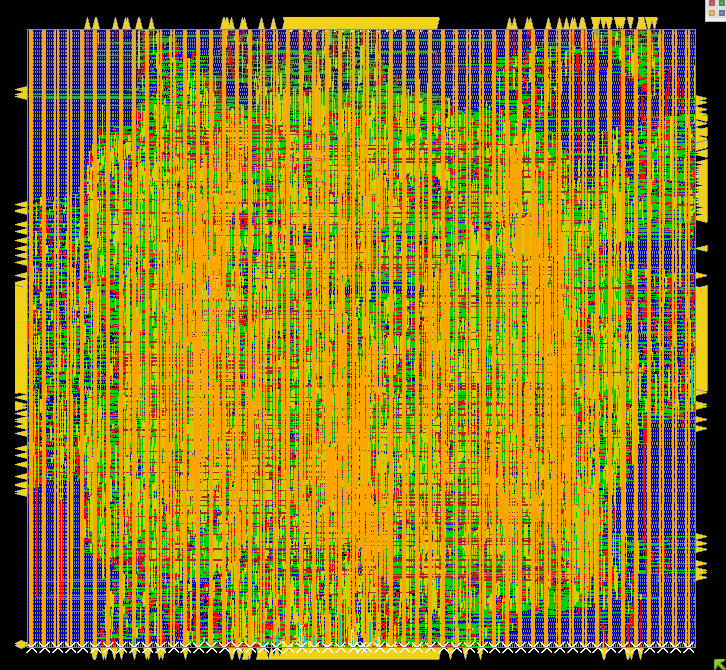}
    \end{subfigure}%
    ~ 
    \begin{subfigure}
        \centering
        \includegraphics[height=1.4in]{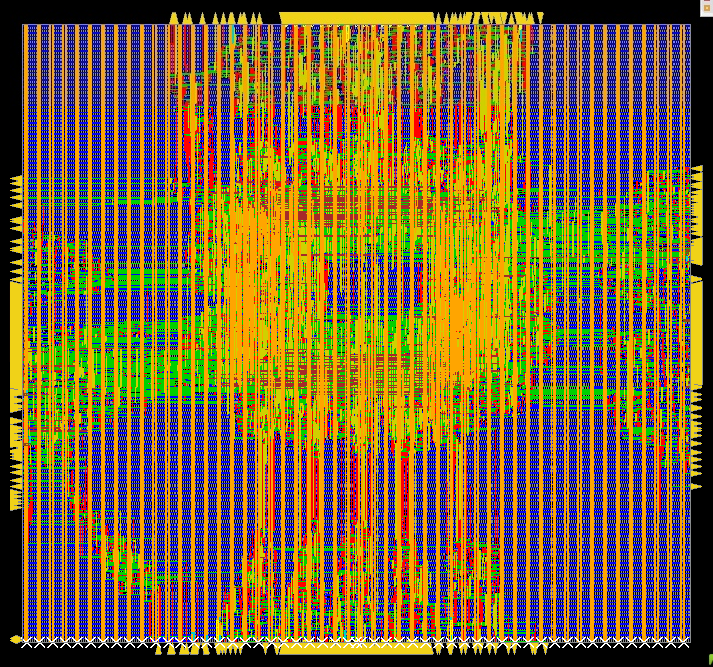}
    \end{subfigure}%
    ~ 
    \caption{Layout plots for CMAC (left) and PCU (right) for an INT4-based $16 \times 4$ array in 45nm CMOS. Note the significant reduction in logic complexity for the latter for the same floorplan size, indicating less area utilization.}
    \label{fig:pnr}
\end{figure}

Place-and-route analysis for INT4-based CMAC and PCU units with $16 \times 4$ array is conducted using 45nm CMOS, maintaining 70\% floorplan utilization for a fair comparison. As shown in Table \ref{tab:pnr}, PCU reduces total area by 53\% and power by 44\% compared to binary multipliers. This confirms the potential of temporal-unary designs for low-precision edge-based DLAs. The area utilization for both designs is illustrated in Fig. \ref{fig:pnr}, which depicts the decrease in area utilization of the PCU design compared to the CMAC unit.

\begin{table}[t!]
\caption{Post-place-and-route results for CMAC and PCU units with a 16x4 array for 45nm CMOS. The total area is in mm$^2$, and the total power is in mW. }
\centering
\resizebox{0.7\columnwidth}{!}{%
\begin{tabular}{|l|l|l|}
\hline
\textbf{Design} &
  \textbf{\begin{tabular}[c]{@{}l@{}}NanGate45\\ Total Area \end{tabular}} &
  \textbf{\begin{tabular}[c]{@{}l@{}}NanGate45\\ Total Power \end{tabular}} \\ \hline
\textbf{CMAC Core} & 0.0361
   & 10.7013\\ \hline
\textbf{\name} & 0.0168
   & 6.1146 \\ \hline
\end{tabular}%
}

\label{tab:pnr}
\end{table}

\begin{figure}
    \begin{subfigure}
    \centering
        \includegraphics[width=0.47\columnwidth,height=3.5cm]{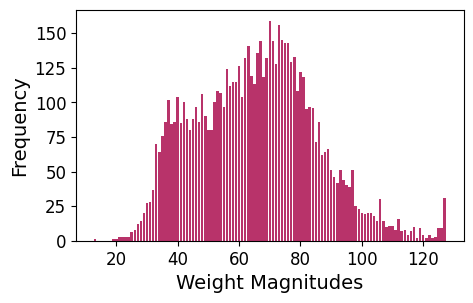}
    \end{subfigure}%
    ~ 
    \begin{subfigure}
    \centering
    \centering
        \includegraphics[width=0.47\columnwidth,height=3.5cm]{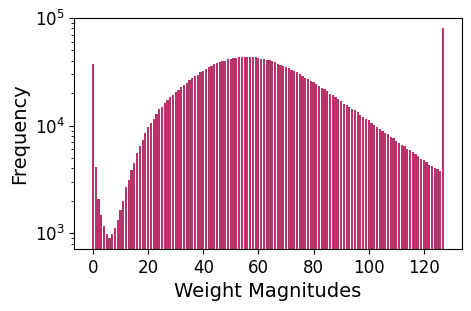}
    \end{subfigure}%
    ~ 
    \caption{Weight magnitude profiling across the convolution layer weights of MobileNetV2 (left) and ResNeXt101 (right) with max pool of $16 \times 16$.}
    \label{fig:weight_dist}
\end{figure}

\begin{figure}
    \begin{subfigure}
    \centering
        \includegraphics[width=0.47\columnwidth,height=3.5cm]{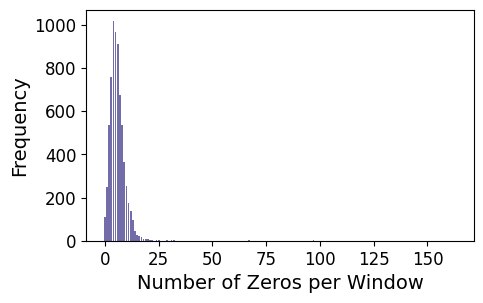}
    \end{subfigure}%
    ~ 
    \begin{subfigure}
    \centering
        \includegraphics[width=0.47\columnwidth,height=3.5cm]{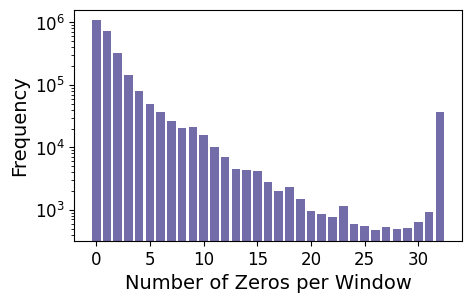}
    \end{subfigure}%
    ~ 
    \caption{Sparsity profiling across the convolution layer weights of MobileNetV2 (left) and ResNeXt101 (right) with tile size of $16 \times 16$.}
    \label{fig:sparsity_dist}
\end{figure}

\subsection{Energy Evaluation with Workload-Dependent Latency}

\name's \textit{2s}-unary encoding \cite{vellaisamy2023tubgemm} allows it to leverage dynamic value sparsity, improving energy efficiency by reducing compute latency. Higher sparsity in the unary encoded signal results in lower latency. We profiled two DNN workloads, MobileNetV2 and ResNeXt101, using max-pooling over $16 \times 16$ tiles of convolution layer weights. Fig. \ref{fig:weight_dist} shows the frequency of weight values (0-128) which correlate with compute latency in a $16 \times 16$ array. The area under the curve normalized by the total sum of frequencies provides the average workload-dependent latency. Using this methodology, MobileNetV2 incurs 33 cycles, and ResNeXt101 incurs 31 cycles, on average, using \textit{2s}-unary encoding. Note that these latencies are almost halved compared to the worst-case INT8 latency of 64 cycles. Using these workload-dependent cycle counts and the 4ns clock period (250 MHz) along with INT8 power consumption values for $16 \times 16$ array (Fig. \ref{fig:pe_16_area_power}), application-specific energy is derived. For the binary-based array, the energy consumption is 15pJ. For \textit{tub}-based array, it amounts to 187pJ and 176pJ for MobilNetV2 and ResNeXt101, respectively, due to the higher compute cycle counts. Although the higher cycle count results in lower energy efficiency for \name for INT8, there is potential to reduce this gap by leveraging zero-value weights to disable the corresponding PE compute. From zero-value weight profiling per $16 \times 16$ tile size in Fig. \ref{fig:sparsity_dist}, the average number of silent PEs for MobileNetV2 is 6 (250 active PEs) and 2 for ResNeXt101. The above energy calculation assumes that all 256 PEs in the tile is active for 31 and 33 cycles, which is an overestimate.

The energy overhead is reduced for lower precision. With INT4, the worst case latency is 4 cycles for tub multipliers with twos-unary encoding.
This results in 7.48pJ for the binary PE array, and 17.76pJ for \textit{tub}-based PE array, indicating energy-gap decrease from 11.7x (INT8) to 2.3x (INT4).

\subsection{Iso-Area Throughput Improvements}
\begin{figure}[h]
\centering
\includegraphics[width=0.5\textwidth]{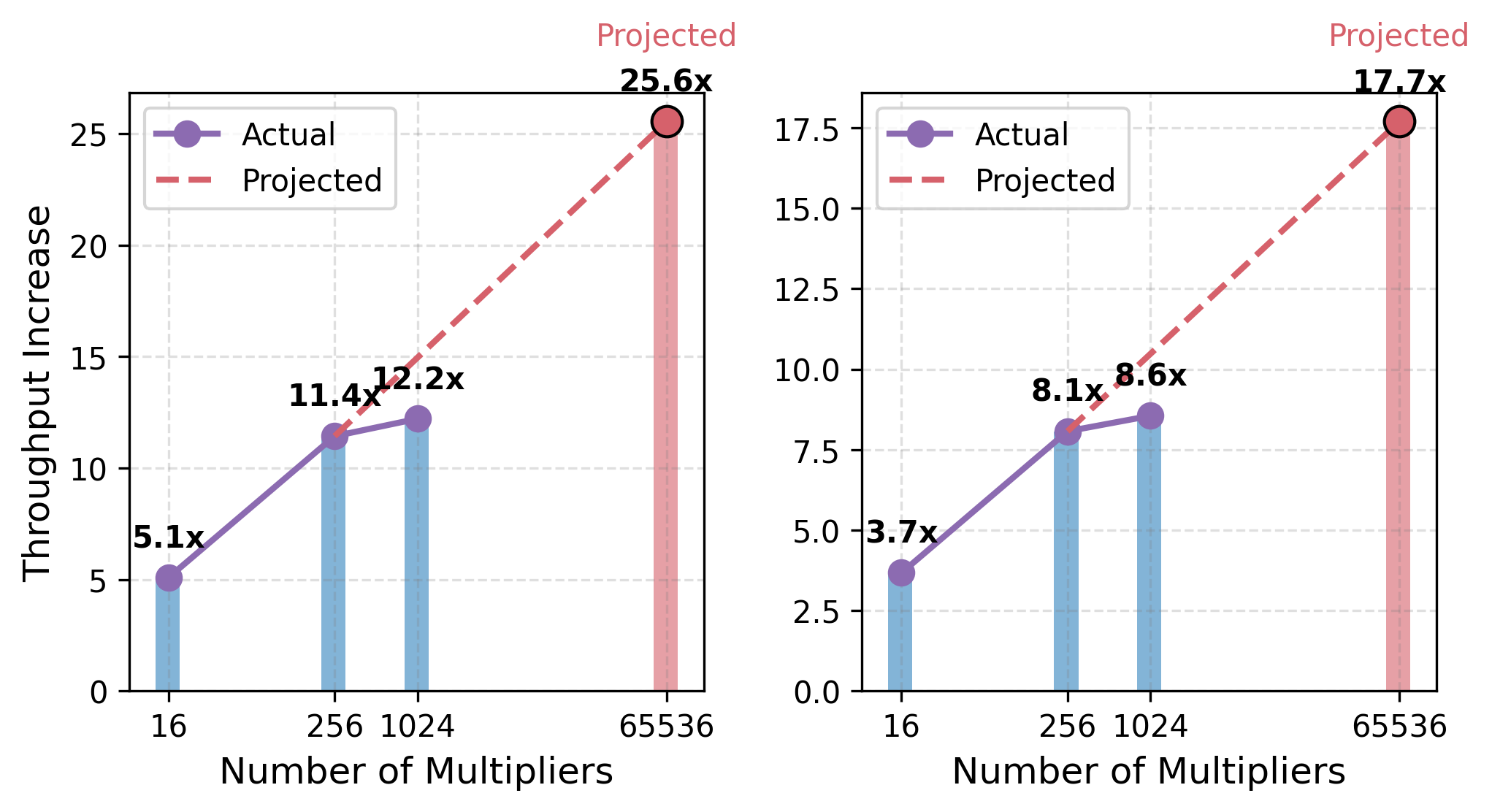}
\caption{Iso-area throughput improvements for a single PE cell ($k=1$) across varying number of multipliers for INT8 (left) and INT4 (right) precisions, assuming same $m$ cycles of latency incurred. Red dotted trend lines project iso-area throughput improvements based on area scaling from Table \ref{tab:pe_area}.}
\label{fig:throughput}
\end{figure}

While the \name design incurs multiple cycles for convolution due to the incorporation of the \textit{tub} PEs, throughput improvements can transcend the latency increase. In NVDLA's binary PE array, $k$ partial sums are produced per cycle, whereas the \textit{tub}-based array in \name requires $m_{k \times n}$ cycles, where $m_{k \times n}$ represents half of the largest weight magnitude in the array (2s-unary encoding). INT8 analysis shows 16 binary PE cells consume 0.09 µm² (from Fig. \ref{fig:pe_16_area_power}), whereas 80 tub PE cells fit in the same area (0.018 µm² for 16 PE cells), yielding a 5x throughput improvement. Similarly, INT4 shows 4x iso-area throughput boost for the same array size. For a single PE cell, scaling across different $n$ number of multipliers (reported in Table \ref{tab:pe_area}), the iso-area throughput improvements are illustrated in Fig. \ref{fig:throughput} for both INT8 and INT4 precisions (assuming the same $m$ cycles to generate one partial product). From the area scaling estimates, we can further project iso-area throughput improvements for a quadratic increase from $n=256$ to $n=65536$ multipliers. The throughput increases by as much as 26x and 18x for INT8 and INT4 precisions, respectively. Hence, based on these observations, we expect iso-area throughput gains to scale even further with increasing array sizes, compensating for the increased compute cycles thus further improving the practical feasibility of \name.

\section{Conclusions and Future Work}
\label{sec:conc}

In this paper, we introduced \name, a temporal-unary-binary hybrid convolution engine optimized for seamless integration into modern edge DLAs. By extending unary computing beyond GEMM-level optimizations to support full convolution operations, \name demonstrates substantial improvement in area and power efficiency. By demonstrating 5x and 4x iso-area throughput improvements for INT8 and INT4 precisions, respectively, for a $16 \times 16$ PE array, we have established a pathway to significantly enhance the performance-per-unit cost ratio of edge AI accelerators. In the future, we would like to extend the work towards unary-based compute architectures targeted towards ultra-low precision quantized large language models (LLMs). Additionally, we aim to explore custom dataflows and compiler optimizations that further reduce latency and energy consumption, enhancing the practicality of unary computing in AI hardware.

\bibliographystyle{IEEEtran}
\bibliography{refs}
\end{document}